# *UNSCT-HRNet: Modeling Anatomical Uncertainty for Landmark Detection in Total Hip Arthroplasty*

Jiaxin Wan[a], Lin Liu[a], Haoran Wang[b], Liangwei Li[a], Wei Li[b]
Shuheng Kou[a], Runtian Li[a], Jiayi Tang[a], Juanxiu Liu[a], Jing Zhang[a], Xiaohui Du[a], Ruqian Hao[a, *]

[a.] *School of Optoelectronic Science and Engineering, University of Electronic Science and Technology of China, Chengdu,611731, China*

[b.] *The Sixth People's Hospital of Chengdu, Chengdu, 610051, China*

# *Abstract*

Total hip arthroplasty (THA) relies on accurate landmark detection from radiographic images, but unstructured data caused by irregular patient postures or occluded anatomical markers pose significant challenges for existing methods. To address this, we propose UNSCT-HRNet (Unstructured CT - High-Resolution Net), a deep learning-based framework that integrates a Spatial Relationship Fusion (SRF) module and an Uncertainty Estimation (UE) module. The SRF module, utilizing coordinate convolution and polarized attention, enhances the model's ability to capture complex spatial relationships. Meanwhile, the UE module which based on entropy ensures predictions are anatomically relevant. For unstructured data, the proposed method can predict landmarks without relying on the fixed number of points, which shows higher accuracy and better robustness comparing with the existing methods. Our UNSCT-HRNet demonstrates over a 60% improvement across multiple metrics in unstructured data. The experimental results also reveal that our approach maintains good performance on the structured dataset. Overall, the proposed UNSCT-HRNet has the potential to be used as a new reliable, automated solution for THA surgical planning and postoperative monitoring.

## 1. Introduction

Total hip arthroplasty (THA)[1-3] is a widely performed surgical procedure, and the procedural success rate heavily depends on the accuracy of the key landmark annotations on the assessment of clinical parameters after accurate annotation of landmarks on the patient's anteroposterior pelvic radiographs[4]. In real-world medical applications, anteroposterior pelvic radiographs often exhibit significant variability and distortion[5], particularly in cases where the patient's condition causes occlusion of important landmarks. For example, hip arthritis can obliterate the joint space[6], femoral head necrosis complicates the identification of landmarks[7], and femoral neck fractures further exacerbate the challenges in landmark detection[8, 9]. Additionally, improper standing posture due to pain can result in further change of landmarks[10, 11], making accurate annotation even more difficult. As a result, radiographic images in practical medical scenarios frequently present in an unstructured form[12], meaning the number of correctly identifiable landmarks does not align with the expected or conventional number. This inconsistency poses significant difficulties for the annotation process.

Currently, landmark annotation is performed either manually or through automated methods based on deep learning. However, manual annotation is time-consuming and subject to inter-observer variability[13, 14], which can compromise both the accuracy and consistency of the measurements. Existing deep learning based automatic annotation methods may produce incorrect results when landmarks are obscured or distorted due to medical conditions. Because these methods rely on local feature patterns alone, and often overlook the spatial and anatomical relationships between landmarks and their broader anatomical significance. In complex hip radiographs, many anatomical points may share similar local characteristics, leading to inaccurate predictions, particularly in cases of occlusion or pathology. Current methods are constrained by their assumption of a fixed number of landmarks, resulting in poor performance when the number of identifiable landmarks is fewer than expected in unstructured data.

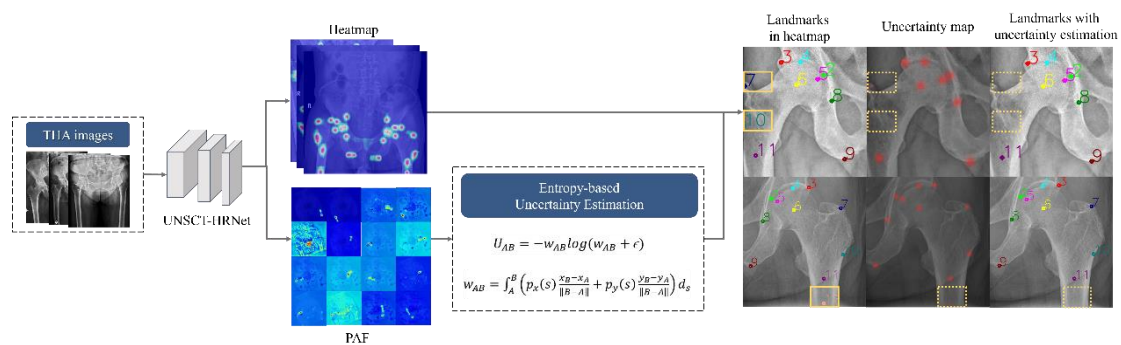

Figure 1 Framework of the UNSCT-HRNet model for THA landmark detection and uncertainty estimation. THA images are first fed into the UNSCT-HRNet model, which generates landmark heatmaps and PAF vectors. The projection weights of the PAF vectors are used as information measures, which are then incorporated into an entropy-based uncertainty formulation. This process yields uncertainty values for each detected landmark, enabling the model to handle unstructured data effectively.

To address these challenges, we propose the UNSCT-HRNet(Unstructured CT - High-Resolution Net) for landmark detection to overcome the limitations posed by unstructured data. UNSCT-HRNet adopts the HRNet(High-Resolution Network)[15] architecture as the backbone and integrates a Spatial Relationship Fusion module (SRF) based on coordinate convolution[16] and polarized attention[17]. This SRF module enhances the model's capability to extract and encode complex spatial relationships from images, enabling more effective capture and comprehension of the spatial configurations of anatomical landmarks, as well as the intricate relationships between them, incorporating these into the feature.

Additionally, we propose an Uncertainty Estimation (UE) module to quantify confidence for each prediction, especially in unstructured data. As shown in Figure 1, The UE module leverages entropy to model uncertainty[18, 19] and utilizes the part affinity fields[20] (PAF) vector to model anatomical relationships between landmarks, which encode anatomical relationships as vector fields. The PAF vector is interpreted as informative data, to assesses the degree of consistency between predicted landmarks, offering a measure of the confidence level in each prediction. By jointly regressing landmark heatmaps and PAF, the model estimates both landmark locations and their directional relationships. The PAF weights quantify uncertainty, helping the model suppress erroneous predictions, particularly when landmarks are occluded or distorted.

In this work, we introduce a robust solution for unstructured radiographic data processing, which referred to as UNSCT-HRNet. UNSCT-HRNet improves the clinical reliability of landmark detection, especially in scenarios where conventional methods falter due to occluded or distorted landmarks. Notably, there is no existing literature reporting the use of uncertainty estimation in medical imaging for landmark detection. This advancement offers critical insights for preoperative planning and postoperative monitoring in THA, ensuring precise evaluations even in complex, unstructured clinical settings.

## *2. Related Work*

In recent years, the application of deep learning in the medical imaging field has continuously made huge breakthroughs. With respect to the THA, deep learning technology has been widely applied in the detection of prosthesis dislocation risk, identification of prosthesis models, and the assessment of various prosthetic parameters[21-23].

Mulford K L et al.[24] used a U-Net architecture with residual blocks to segment 22 different structures including landmarks, but the point structure mask generated by U-Net provides an approximate range of the landmark, rather than precise pixel-level coordinates. Rouzrokh P et al.[25] utilized two U-Nets to perform segmentation of the acetabular component and both ischial tuberosities on anteroposterior pelvis radiographs and cross-table lateral hip radiographs. The longitudinal axis of the acetabular component is determined by the outermost and innermost points of the acetabular mask, while the trans-ischial tuberosity line is defined by the lower two points of the ischial mask. The Inclination angle and Anteversion angle are subsequently calculated. However, due to the inability to directly determine the certain landmarks based on the spatial position of the segmented structure, the generalizability of this method is limited. Moon K R et al.[26] employed the ResNet50 network for regression to detect the rough landmark' region of interest within an original radiographs. Subsequently, the region of interest is inputted back into the ResNet50 network again to extract a more precise coordinate position. This dual-stage pelvic landmark detection network identifies 16 landmark locations and computes 10 radiographic parameters based on these landmarks. Tsung-Wei Tseng et al.[27] proposed the BKNet for the identification of 20 landmarks and calculation of 10 parameters in postoperative anteroposterior pelvic radiographs, with localization errors of each landmark being within 1 cm. Wei Yang et al.[28] utilized a CNN with skip connect to detect specific points in the acetabulum, subchondral sclerosis, teardrop, and femoral head center, achieving nearly 93% accuracy in landmark detection within 3mm and an error of 6° in angle measurement.

While existing methods show promise in detecting landmarks and computing parameters, they primarily focus on improving detection accuracy for structured data and often overlook the challenges posed by unstructured data. A key limitation is the lack of comprehensive research that addresses unstructured data in landmark predictions. This study proposes the UNSCT-HRNet to handle unstructured radiographic data, effectively predicting landmarks in images with variable and non-fixed landmarks. Additionally, UNSCT-HRNet incorporates a UE module based on entropy, modeling the anatomical relationships of landmarks with PAF and allowing for the quantification of uncertainty in landmark predictions. This capability not only improves robustness in complex medical scenarios but also enhances the accuracy of clinically significant parameter calculations from radiographs of THA patients, providing a more reliable and comprehensive digital assessment tool.

## 3. Method

### 3.1 Task Conversion

In this study, we focus on the task of accurately detecting anatomical landmarks from radiographic images, which is crucial for evaluating the outcomes of THA. The input to UNSCT-HRNet consists of high-resolution pelvic radiographs, and the desired output is a set of precise landmark coordinates that facilitate clinical assessments.

We employ HRNet as the backbone network with a multi-scale architecture to maintain high-resolution representations for accurate anatomical landmark detection. The SRF module based on polarized attention and coordinate convolution is introduced to enhance the extraction of spatial relationship features. Landmarks are regressed through heatmaps, with Adaptive Wing Loss[29] used to balance the contributions of positive and negative samples. To handle unstructured data and estimate prediction uncertainty, we also regress PAF to quantify the uncertainty of landmark predictions.

## 3.2 Model Architecture

### 3.2.1 High-Resolution Network

In this study, we leverage the HRNet with a multi-scale branch architecture as the backbone of our model. The reason of choosing HRNet as backbone is because it has ability to maintain high-resolution features throughout the processing stages, which is crucial for the precise detection of landmarks in radiographic images and accurate part affinity field regression. The architecture's parallel branches capture diverse levels of detail, allowing for a comprehensive representation that enhances the prediction accuracy of landmarks and part affinity fields.

To further improve the model's performance, we have integrated two key components into the HRNet architecture: a SRF module and a UE module. These enhancements will be discussed in the subsequent sections. The model structure is shown in Figure 2.

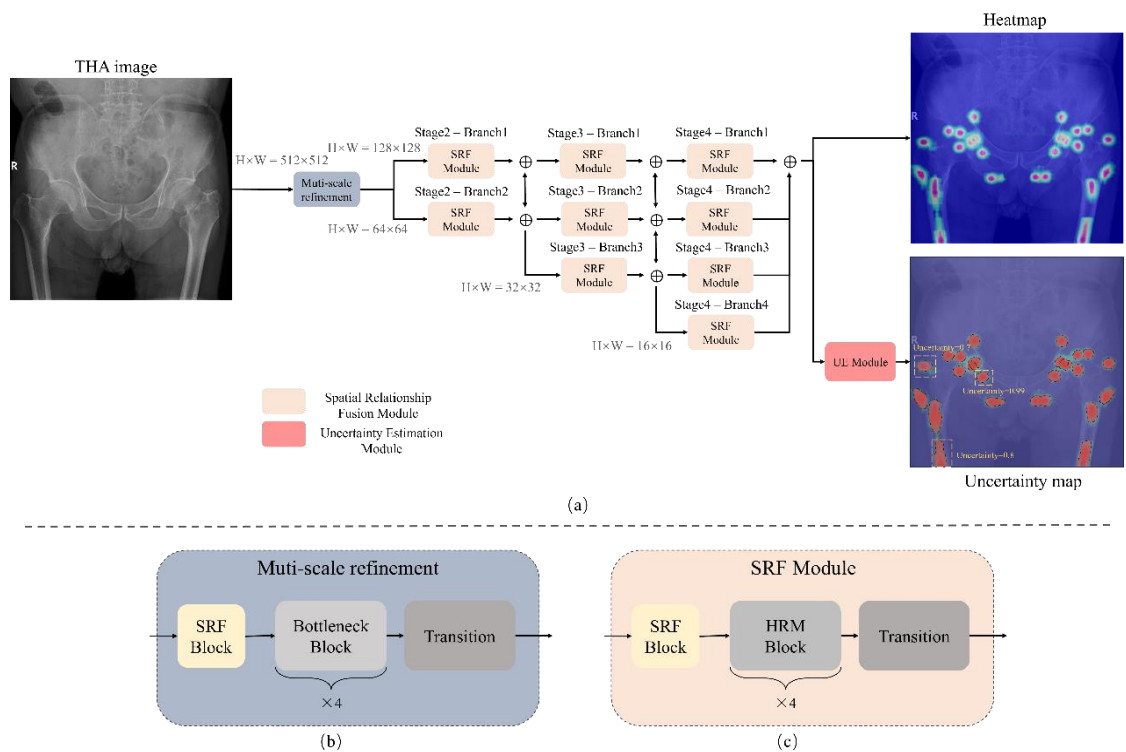

Figure 2 Architecture of the proposed model for THA landmark detection. (a) The overall process of the model, which takes an input image and processes it through multiple layers composed of SRF module. The final stage is an uncertainty estimation module that generates uncertainty map, visualized on the right. (b) Detailed structure of Muti-scale refinement, which includes a SRF block, four bottleneck blocks, and a transition block to progressively refine features. (c) Detailed structure of a SRF Module, which consists of a SRF block, four HRM blocks, and a transition block, repeated to further enhance feature extraction at different levels.

### 3.2.2 Spatial Relationship Fusion Module

The annotation of anatomical landmarks in THA images usually has spatial dependencies. Therefore, it is necessary to be able to capture the spatial relationship features between landmarks in landmark detection[30]. To enhance the network's capability to capture the spatial relationships among the landmarks in THA images, we incorporate heatmap and coordinate information for all landmarks in each branch. This information is fused using 1×1 convolutions, allowing the model to integrate feature representations effectively. Furthermore, we employ polarized spatial and channel attention to capture the relationships between landmarks. The SRF module combining coordinate convolution and polarized attention can better regress the part affinity field created based on anatomical significance. The structure of the SRF module is shown in Figure 3.

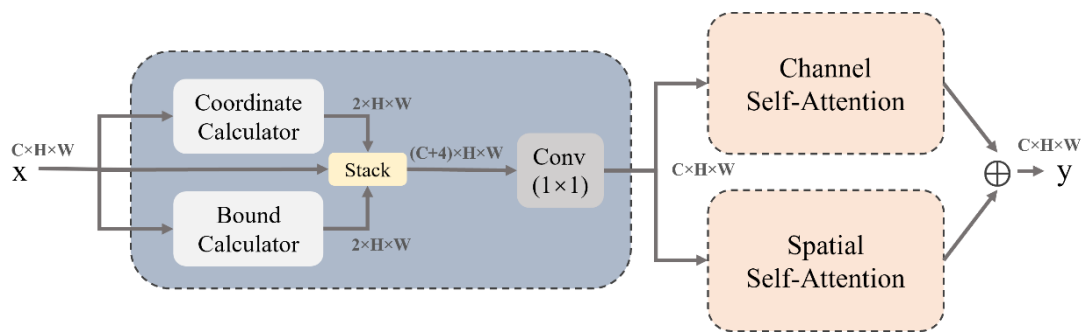

Figure 3 The SRF module enhances spatial relationship learning among landmarks in THA images. Coordinate and boundary information is calculated and stacked with the feature map, followed by a 1×1 convolution for integrated feature representation. Polarized channel and spatial self-attention then capture spatial and channel correlations among landmarks, facilitating accurate regression of part affinity fields based on anatomical relevance.

### *3.2.3 Uncertainty Estimation Module*

To address unstructured data and quantify the uncertainty in landmark prediction, we introduce a lightweight UE module that utilizes entropy to model uncertainty. In this module, the PAF vectors are treated as a source of information content, representing the anatomical relationships between landmarks. This module simultaneously regresses the heatmap of the landmarks and predicts the PAF vector representing the anatomical relationship between the landmarks.

The concept of PAF was originally employed to model the associations between different body parts by representing the anatomical relationships as vector fields[20]. Specifically, PAF encode the location and orientation of limbs as vectors connecting joints, allowing the model to estimate both the presence of landmarks and the connections between them in a consistent and coherent manner. This method enables robust pose estimation by linking detected landmarks through their geometric relationships, even in cases of occlusion or crowded scenes. By leveraging these part-to-part relationships, PAF improves the reliability and accuracy of multi-person pose estimation, providing a powerful tool for capturing complex spatial configurations in real-time scenarios.

In our work, PAF is used to model the actual anatomical relationships between landmarks, including the anatomically significant distances and angular relationships between two landmarks. Each PAF vector is constructed with two channels, representing the normalized x and y unit vector values that correspond to the directional relationship between the two landmarks. These anatomical relationships, such as distances and angles, are carefully defined under the guidance of professional clinicians to ensure their clinical relevance and accuracy.

To quantify landmark prediction uncertainty, we constructed an entropy-based uncertainty formulation:

$$U_i = -w_i log(w_i + \varepsilon) \qquad (1)$$

where $U_i$ denotes the uncertainty quantification of the i-th landmark, and $w_i$ represents the projection weight of the corresponding PAF vector along the direction connecting the landmarks. This weight reflects the degree of spatial consistency. The constant ε is included to prevent numerical instability when $w_i$ approaches zero. This formula is based on the principle of information entropy, indicating that as the projection weight $w_i$ increases, uncertainty $U_i$ decreases, indicating stronger consistency with anatomical prior knowledge. Conversely, a smaller projection weight $w_i$ implies greater uncertainty $U_i$, suggesting higher deviation from expected anatomical relations.

In this context, we calculate the coordinates of the predicted landmarks $A = (x_A, y_A)$ and $B = (x_B, y_B)$ from the heatmaps and define the unit direction vector connecting these points as follows:

$$d_{AB} = \frac{B-A}{\|B-A\|} = (\frac{x_B - x_A}{\|B-A\|}, \frac{y_B - y_A}{\|B-A\|}) \qquad (2)$$

For a given location in the PAF field, the vector $p_{AB} = (p_x, p_y)$ represents the PAF vector at that location. The projection weight $w_{AB}$ along the direction vector $d_{AB}$ is then calculated by:

$$w_{AB} = \int_A^B \left( p_x(s)\frac{x_B - x_A}{\|B-A\|} + p_y(s)\frac{y_B - y_A}{\|B-A\|} \right) d_s \qquad (3)$$

This integration provides the accumulated projection weight $w_{AB}$

along the path from *A* to *B*, capturing the spatial consistency of the PAF field along this direction. Substituting $w_{AB}$ into the entropy formula yields the uncertainty $U_{AB}$ for landmarks *A* and *B*:

$$U_{AB} = -w_{AB}\log(w_{AB}+\varepsilon) \tag{4}$$

If the projection weight $w_{AB}$ closely approximates the magnitude of *B-A*, it suggests alignment with the expected anatomical relationship, leading to a lower $U_{AB}$ value. Conversely, if $w_{AB}$ is close to zero, it indicates deviation from expected spatial consistency, increasing $U_{AB}$ and hence, the uncertainty.

Each landmark may correspond to multiple PAF vectors. Therefore, the uncertainty related to the PAF angles for a landmark is derived from a weighted sum of all cumulative PAF projection values associated with that landmark. The predefined PAF vector is shown in Figure 4.

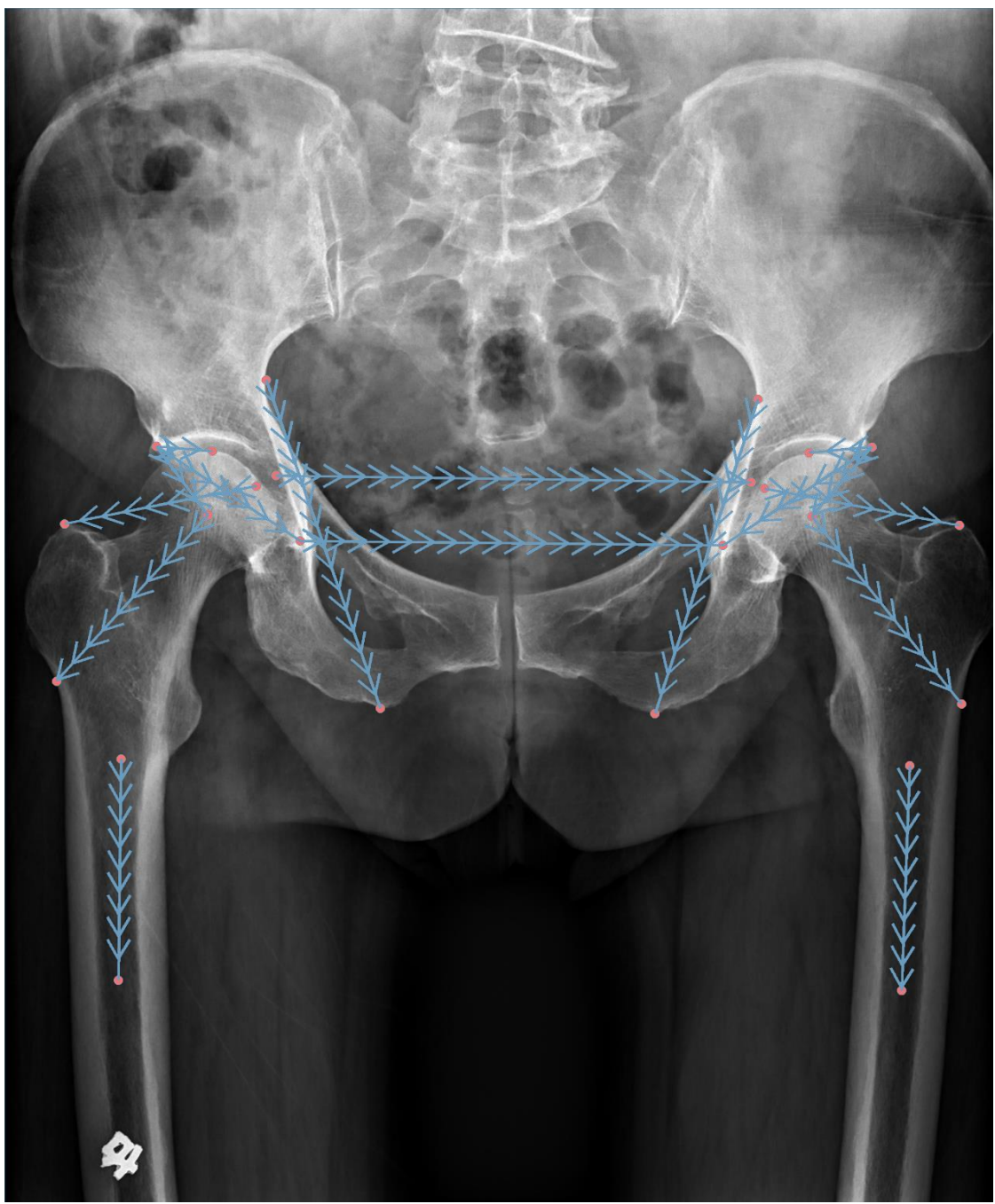

Figure 4 Predefined PAF between anatomical landmarks in THA radiographs. Each PAF vector represents the directional relationship between two landmarks, encoded with two channels for normalized x and y unit vectors. These vectors reflect anatomically significant distances and angular relationships, defined with clinical guidance. The figure illustrates the PAF vectors used to model these relationships.

## 3.3 Loss Function

In this study, the anteroposterior pelvic radiographs we used contains a substantial amount of irrelevant background information. During landmark heatmap regression, positive samples often comprise only a few hundred pixels, while negative samples can amount to tens of thousands of pixels[31]. As a result, employing common loss functions for landmark detection, such as Mean Square Error (MSE), would cause positive samples to be overwhelmed by the vast number of negative samples, leading to poor training performance[32]. To address this issue, we adopted a hybrid loss function approach for model training.

In our landmark heatmap regression, we applied Adaptive Wing Loss, which offers advantages over traditional loss functions like MSE by effectively addressing the imbalance between positive and negative samples. The calculation method of Adaptive Wing Loss is shown in Formula (5).

$$L = \begin{cases} \omega \cdot \log(1 + (\frac{|pred - gt|}{\varepsilon})^{\alpha - gt}) & if\ |pred - gt| \leq \theta \\ A \cdot |pred - gt| - C & if\ |pred - gt| > \theta \end{cases} \quad (5)$$

$$A = \omega \cdot \left(\frac{1}{1+\left(\frac{\theta}{\varepsilon}\right)^{\alpha-gt}}\right) \cdot (\alpha - gt) \cdot \left(\frac{\theta}{\varepsilon}\right)^{\alpha-gt-1} \cdot \frac{1}{\varepsilon} \quad (6)$$

$$C = \theta \cdot A - \omega \cdot \log\left(1+\left(\frac{\theta}{\varepsilon}\right)^{\alpha-gt}\right) \quad (7)$$

Where *A* and *C* are calculated constants, as shown in formulas (6) and (7). Adaptive Wing Loss adapts its focus based on the distribution of samples, allowing for more accurate learning even when positive samples are scarce. To further enhance this balance, we implemented channel and spatial masking, which allows us to strategically weigh the importance of different regions in the heatmap. This masking approach ensures that positive samples are prioritized, ultimately improving the model's performance during training. The masked Adaptive Wing Loss is shown in formula (8).

$$\textit{mask Adaptive Wing Loss} = L \cdot (W \cdot mask + 1) \quad (8)$$

The convergence comparison between Adaptive Wing Loss and the original MSE Loss on landmark heatmap regression is shown in Figure 5. The convergence performance of the Adaptive Wing Loss model is significantly improved.

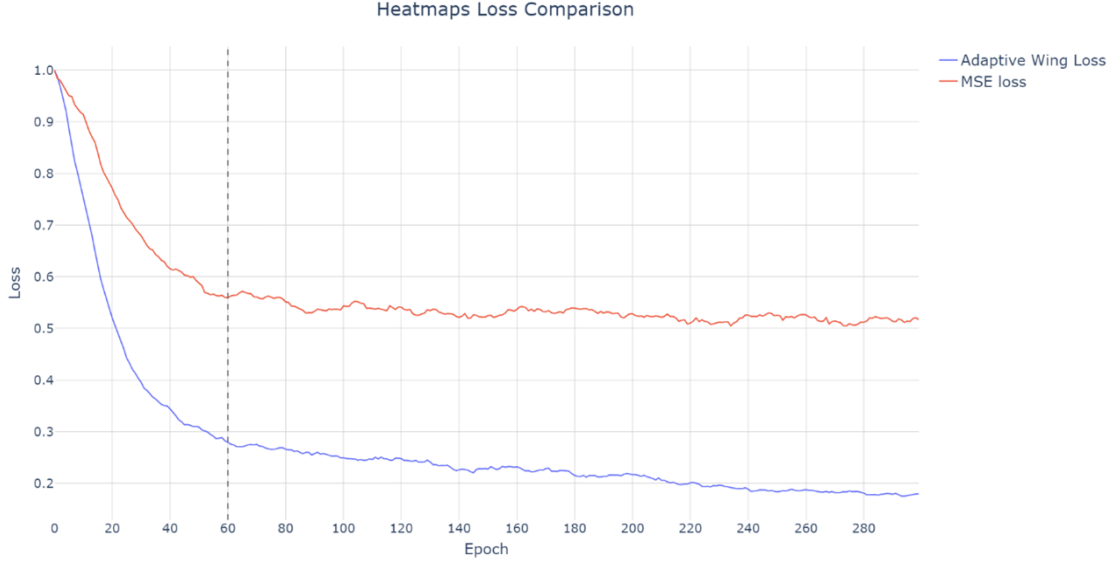

Figure 5 Heatmap loss comparison between Adaptive Wing Loss and MSE Loss. The Adaptive Wing Loss (blue) shows faster convergence and lower loss than MSE Loss (red), indicating better performance in handling imbalanced landmark predictions.

For the PAF regression, we used MSE to compute the loss. The final hybrid loss function is a weighted sum of Adaptive Wing Loss and Mean Square Error.

$$Loss = W_1 L_{mask\ Adaptive\ Wing\ Loss} + W_2 L_{PAF} \qquad (9)$$

## 3.4 Training Procedure

We implemented the training procedure for UNSCT-HRNet using PyTorch 2.0.1, with a CUDA version of 11.7 to leverage GPU acceleration on an NVIDIA RTX 3090 24 GB. The training was conducted over a total of 300 epochs, allowing sufficient time for the model to converge and improve its performance on the landmark detection task. The optimizer of choice was Adam, which adapts the learning rate for each parameter, ensuring efficient convergence.

The learning rate was set to 0.001, which strikes a balance between convergence speed and stability. To enhance the model's robustness and prevent overfitting, we employed a dynamic data augmentation strategy during training, which included techniques such as random flipping and adding noise to the images.

In our experimental design, we trained two versions of the HRNet: one without the UE module and one with it. This comparison allowed us to evaluate the impact of the UE module on the model's ability to handle unstructured data. The model with the UE module incorporates PAF to quantify uncertainty in landmark predictions, enhancing its robustness in scenarios where landmarks may be occluded or distorted.

During each epoch, the model was trained on the training set consisting of 257 images, while validation was performed on a separate set of 53 images. We monitored validation loss and other performance metrics throughout the training to ensure effective learning and to avoid overfitting. By analyzing the performance differences between the two models, we aimed to demonstrate the advantages of integrating the UE module in accurately detecting anatomical landmarks in radiographic images.

## 3.5 Evaluation Metrics

To thoroughly evaluate the performance of our landmark detection model, we utilized several evaluation metrics that assess both the accuracy of the predicted landmarks and their consistency with clinical annotations. These metrics ensure that predictions are reliable for clinical applications, particularly for the evaluation of THA outcomes.

Normalized Mean Error (NME) normalizes the error between predicted and true landmarks, allowing comparison across images of different sizes and resolutions.

$$NME = \frac{1}{N}\sum_{i=1}^{N}\frac{\|p_i - g_i\|}{d} \qquad (10)$$

Where $p_i$ is the predicted landmark, $g_i$ is the ground truth landmark, and d is a normalization factor based on the inter-landmark distance or another reference distance.

Mean Radial Error (MRE) is the average Euclidean distance in millimeters between predicted and true landmarks, indicating model accuracy.

$$MRE = \frac{1}{N}\sum_{i=1}^{N}\sqrt{\left((x_i^{pred} - x_i^{true}) \cdot x_i^{ps}\right)^2 + \left((y_i^{pred} - y_i^{true}) \cdot y_i^{ps}\right)^2} \quad (11)$$

Where N is the number of landmarks, $(x_i^{pred}, y_i^{pred})$ is the predicted position of the i-th landmarks, $(x_i^{true}, y_i^{true})$ is the corresponding true position, $(x_i^{ps}, y_i^{ps})$ is the pixel spacing. This formula provides a quantifiable measure of the model's prediction accuracy, with smaller MRE values indicating higher precision.

We use SDR to evaluate the percentage of predicted landmarks within a 2mm error threshold, indicating how often the predictions meet clinical standards. Additionally, PCC, ICC, and the T-test are employed to assess the consistency between model predictions and clinical annotations, and to determine if there are any statistically significant differences.

These evaluation metrics provide a comprehensive analysis of the model's performance, focusing on accuracy and consistency with clinical standards. They ensure that the model is accurate and reliable in predicting unstructured data, and is suitable for clinical application in THA assessment.

## *4. Experiments and Evaluation*

### *4.1 Data Preparation*

#### *4.1.1 Clinical Parameters*

The evaluation of preoperative angles based on landmarks is crucial for maintaining patient health and is an essential component of THA. Various angle and dimension measurements are used as evaluation annotations. Preoperatively, the Skinner line assesses the presence of a displaced fracture in the femoral neck or greater trochanter. The femoral neck-shaft angle is a key metric for diagnosing coxa vara and coxa valga, while the femoral offset is significant for reconstructing hip joint biomechanics.

The acetabular offset is critical for maintaining proper alignment and function of the hip joint, influencing both range of motion and load distribution. The center-edge angle assesses the relationship between the acetabulum and the femoral head, which is vital for determining stability within the acetabular socket. The acetabular index is pivotal for diagnosing and evaluating congenital hip dislocation, and the Sharp angle aids in diagnosing acetabular dysplasia.

To accurately compute these clinical parameters, specific anatomical landmarks must be detected. As shown in the Figure 6, the identified landmarks correspond to the parameters mentioned above, facilitating a comprehensive analysis of the patient's preoperative condition. The landmarks necessary for these evaluations are detailed in Table 1, with the corresponding parameters listed in Table 2.

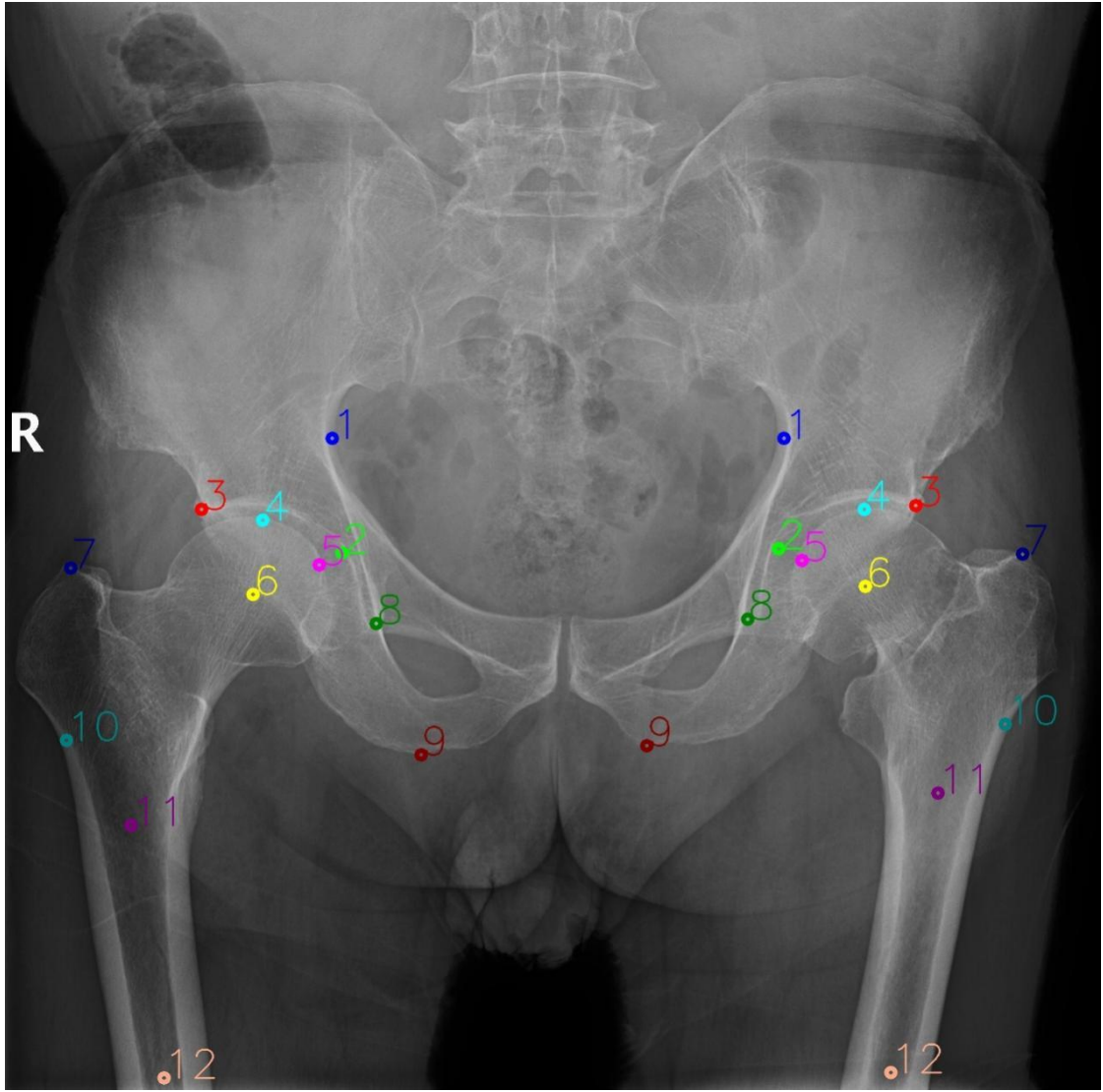

Figure 6 Landmark annotations on THA image. This image displays 24 annotated landmarks representing 12 distinct anatomical categories, with each category marked symmetrically on both the left and right sides of the hip. The landmarks are color-coded and numbered to differentiate each category, facilitating precise localization of anatomical structures relevant for THA analysis.

Table 1. Predicted landmarks

| Landmarks | Index |
|---|---|
| Left- and Right- Innermost point of the ilium | 1 |
| Left- and Right- Center of the Y-shaped cartilage | 2 |
| Left- and Right- Upper edge of the acetabular surface | 3 |
| Left- and Right- Lower edge of the acetabular surface | 4 |
| Left- and Right- Fovea of the ligamentum teres | 5 |
| Left- and Right- Center of the femoral head | 6 |
| Left- and Right- Tip of the greater trochanter | 7 |
| Left- and Right- Lower edge of the teardrop | 8 |
| Left- and Right- Bottom of the ischium | 9 |
| Left- and Right- Distal midpoint of the femoral neck | 10 |
| Left- and Right- Proximal midpoint of the femoral shaft | 11 |
| Left- and Right- Distal midpoint of the femoral shaft | 12 |

Table 2. Clinical parameters and landmarks required

| Clinical Parameters | Landmarks used |
|---|---|
| Left- and Right- Skinner line | 5, 7, 11, 12 |
| Left- and Right- femoral neck-shaft angle | 6, 10, 11, 12 |
| Left- and Right- femoral offset | 6, 11, 12 |
| Left- and Right- acetabular offset | 6, 8 |
| Left- and Right- center-edge angle | 3, 6 |
| Left- and Right- acetabular index | 2, 3, 4 |
| Left- and Right- Sharp angle | 3, 8 |
| Left- and Right- Kohler line | 1, 9 |

### *4.1.2 Dataset*

In this study, preoperative anteroposterior pelvic radiographs were collected from August 2017 to January 2023. A total of 284 patients were included, with 204 having available images. Due to extended treatment cycles, a single patient may have multiple radiographs. Rigorous selection criteria were implemented to ensure data accuracy, resulting in 310 preoperative radiographs from Chengdu No.6 People's Hospital. All identifiable information was removed to protect patient privacy.

The original DICOM files were converted to JPEG images while maintaining their resolution. We employed a dynamic data augmentation strategy during training, which included techniques such as flipping and noise addition, without increasing the total number of training and validation images. The dataset consists of 257 training images and 53 validation images, ensuring no overlap of images from the same patient between the two sets. Additionally, normalization was applied to standardize the data for better model performance. The 257 training images contain 32 unstructured data and 225 structured data. The 53 validation images contain 18 unstructured data and 35 structured data. Most of the unstructured data are missing at the 10th and 12th points. Figure 7 shows common unstructured data in the dataset.

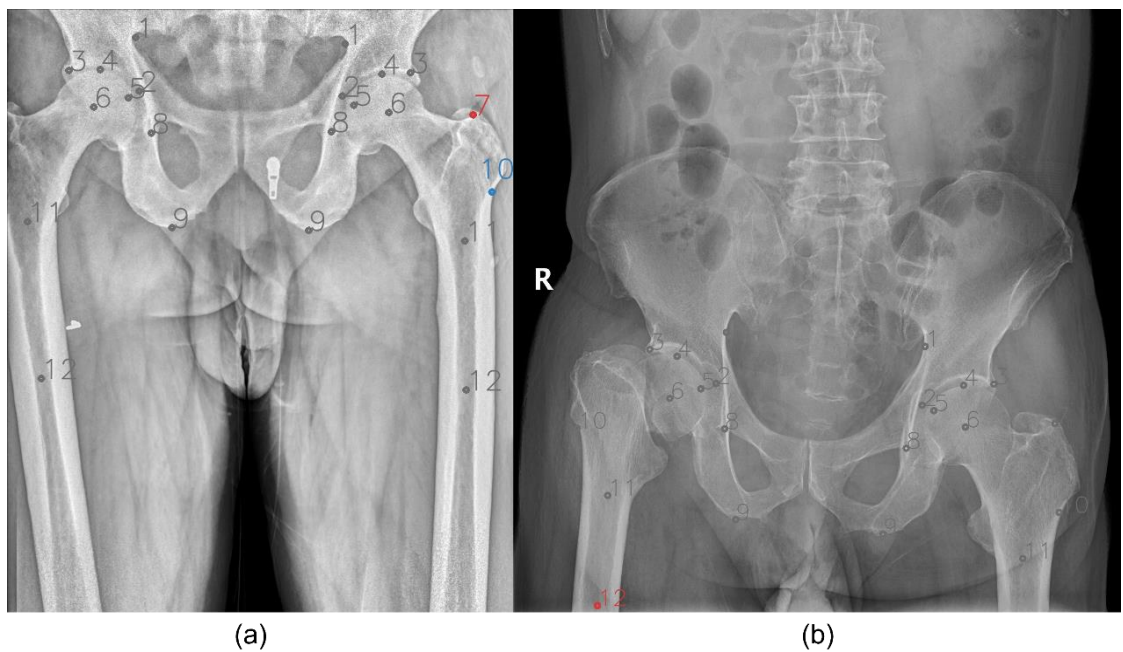

(a)  (b)

Figure 7 (a) The 7th and 10th landmarks on the left side of the THA image are obscured due to occlusion. (b) The 12th landmark on the right side of the THA image is missing due to the patient's atypical posture. Normally, the 12th landmark should be located at the Distal midpoint of the femoral shaft.

## 4.2  Heatmap and PAF regression

To illustrate the model's learning process, Figure 8 present the visualizations of the landmark heatmap regression and PAF regression during training. The heatmap regression highlights the model's ability to localize individual landmarks with high precision, while the PAF regression captures the anatomical relationships between these landmarks, providing additional structural context. These visualizations clearly demonstrate how the model learns to accurately predict both the positions of the landmarks and their geometric relationships, even in challenging scenarios involving unstructured data.

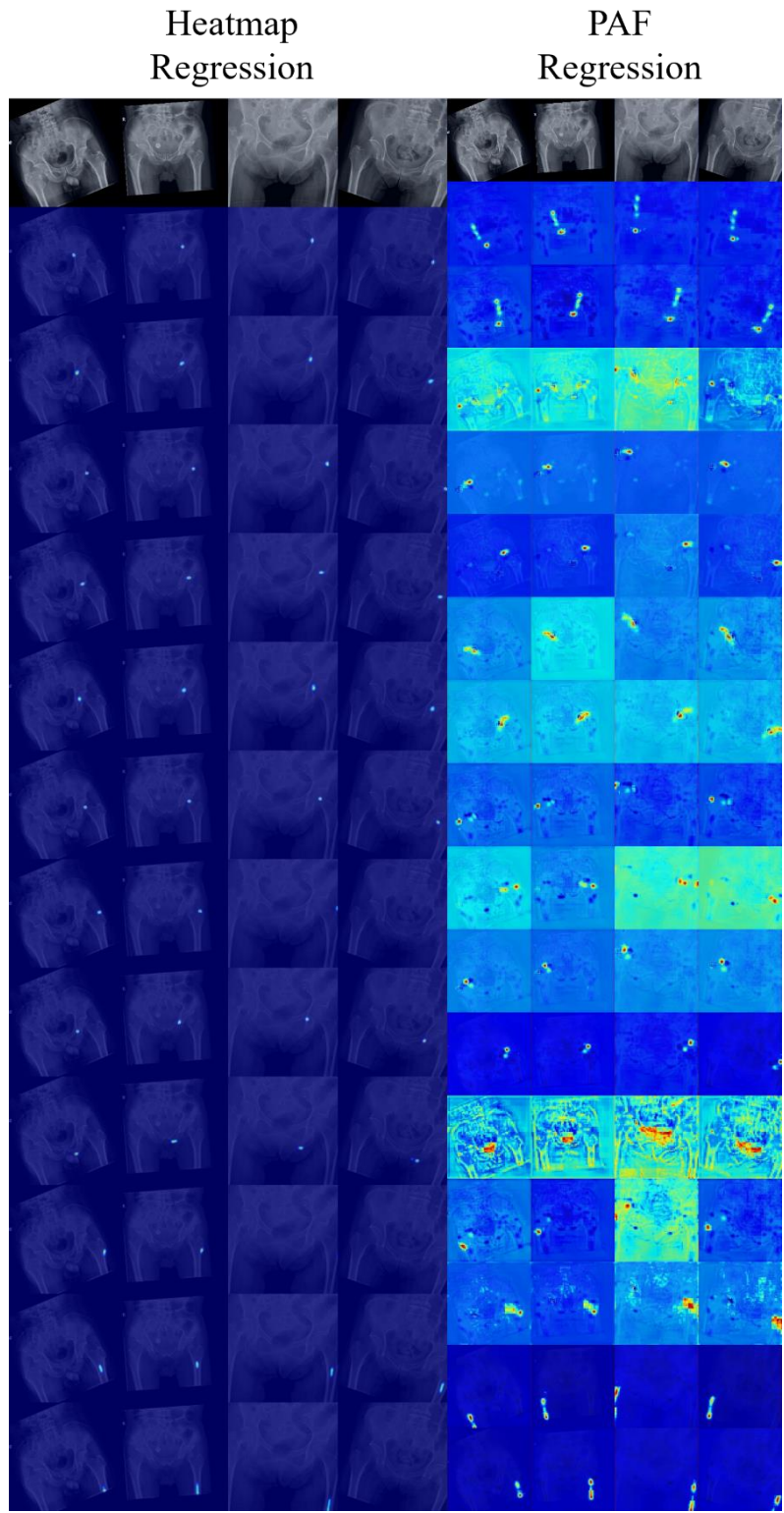

Figure 8 Landmark detection heatmap regression and PAF regression. Each subfigure in the rows of the heatmap regression represents the heatmap for a specific landmark, while each subfigure in the rows of the PAF regression represents the illustration of a specific PAF vector. This visualizes both the landmark locations and the directional relationships between them through PAF.

## 4.3 Performance

### 4.3.1 SRF module integrates spatial relationship features

The SRF module significantly enhances HRNet's sensitivity to spatial relationships between landmarks, improving its ability to capture complex anatomical structures. As shown in Table 3, we conducted ablation experiments on the SRF module across the entire validation set, with the UE module disabled, and observed that the addition of the SRF module yields substantial improvements across all metrics. Specifically, NME and MRE decrease by 40.00% and 40.36%, respectively, indicating enhanced localization accuracy. PCC and ICC scores increase by 12.41% and 12.57%, respectively, reflecting stronger consistency in landmark prediction. Furthermore, SDR increases by 25.45%, suggesting a higher success rate in accurate landmark detection. These results substantiate the effectiveness of the SRF module in augmenting HRNet's spatial reasoning capabilities for improved landmark sensitivity.

Table 3 SRF module ablation experiment

| Model | NME | MRE | SDR | PCC | ICC | T-Test |
|---|---|---|---|---|---|---|
| HRNet | 0.145 | 12.479 | 0.617 | 0.846 | 0.843 | 0.331 |
| HRNet+SRF module | 0.087 | 7.442 | 0.774 | 0.951 | 0.949 | 0.403 |

*4.3.2 UE module suppresses non- anatomically predictions*

To validate the effectiveness of the proposed UE module for unstructured CT images, we applied the UE module to three different landmark detection architectures: HRNet, HigherHRNet, and HourglassNet. Models without the UE module are denoted as ori-, while models incorporating the UE module are denoted as UNSCT-.

We evaluated the model performance on two subsets: structured data and unstructured data, to fully demonstrate the advantages of our method in processing unstructured data. Table 4 presents the performance of each ori-model and UNSCT-model on structured data, indicating that the performance of each UNSCT-model is comparable to that of the ori-model. In Table 5, the UE module significantly enhanced performance across all models on unstructured data. For NME, UNSCT-HRNet, UNSCT-HigherHRNet, and UNSCT-HourglassNet showed reductions of 74.62%, 67.30%, and 31.02%, respectively, indicating notable improvements in landmark localization accuracy. In terms of MRE, reductions were similarly significant, with improvements of 79.06% for UNSCT-HRNet, 70.38% for UNSCT-HigherHRNet, and 34.96% for UNSCT-HourglassNet, highlighting the module's impact on reducing mean localization error. Furthermore, PCC and ICC showed consistent gains across all models; UNSCT-HRNet exhibited increases of 46.03% and 46.31%, UNSCT-HigherHRNet demonstrated improvements of 43.01% and 43.76%, and UNSCT-HourglassNet showed respective increases of 19.59% and 20.14%, confirming the enhanced reliability and consistency of landmark predictions facilitated by the UE module. Among all architectures, the UE module has the greatest performance improvement on UNSCT-HRNet.

Table 4 structured data performance

| Model | NME | MRE | SDR | PCC | ICC | T-Test |
|---|---|---|---|---|---|---|
| Ori-HRNet | 0.085 | 7.228 | 0.915 | 0.967 | 0.964 | 0.322 |
| UNSCT-HRNet | 0.073 | 6.149 | **0.943** | 0.975 | 0.972 | **0.391** |
| Ori-HigherHRNet | 0.087 | 6.245 | 0.833 | 0.967 | 0.964 | 0.337 |
| UNSCT-HigherHRNet | 0.073 | 6.161 | **0.943** | 0.975 | **0.973** | 0.387 |
| Ori-HourglassNet | **0.068** | **5.637** | **0.943** | **0.976** | **0.973** | 0.290 |
| UNSCT-HourglassNet | 0.098 | 8.161 | 0.800 | 0.919 | 0.916 | 0.281 |

Table 5 unstructured data performance

| Model | NME | MRE | SDR | PCC | ICC | T-Test |
|---|---|---|---|---|---|---|
| Ori-HRNet | 0.402 | 39.189 | 0.111 | 0.630 | 0.626 | 0.385 |
| UNSCT-HRNet | **0.102** | **8.206** | **0.688** | **0.920** | 0.916 | 0.445 |
| Ori-HigherHRNet | 0.318 | 28.316 | 0.111 | 0.643 | 0.638 | 0.364 |
| UNSCT-HigherHRNet | 0.104 | 8.390 | **0.688** | 0.919 | **0.917** | **0.446** |
| Ori-HourglassNet | 0.187 | 16.485 | 0.125 | 0.725 | 0.720 | 0.335 |
| UNSCT-HourglassNet | 0.129 | 10.714 | 0.625 | 0.867 | 0.865 | 0.368 |

Figure 9 shows the uncertainty map generated by the UE module, as well as the comparison of the marker output without and with the UE module. The uncertainty map indicates the anatomical certainty of the landmarks predicted by the heat map, and the brightness indicates the degree of certainty. The brighter the marker, the more anatomically significant it is. UE module can correctly suppress the output of markers that have no anatomically significance, ensuring the accuracy and robustness of the results.

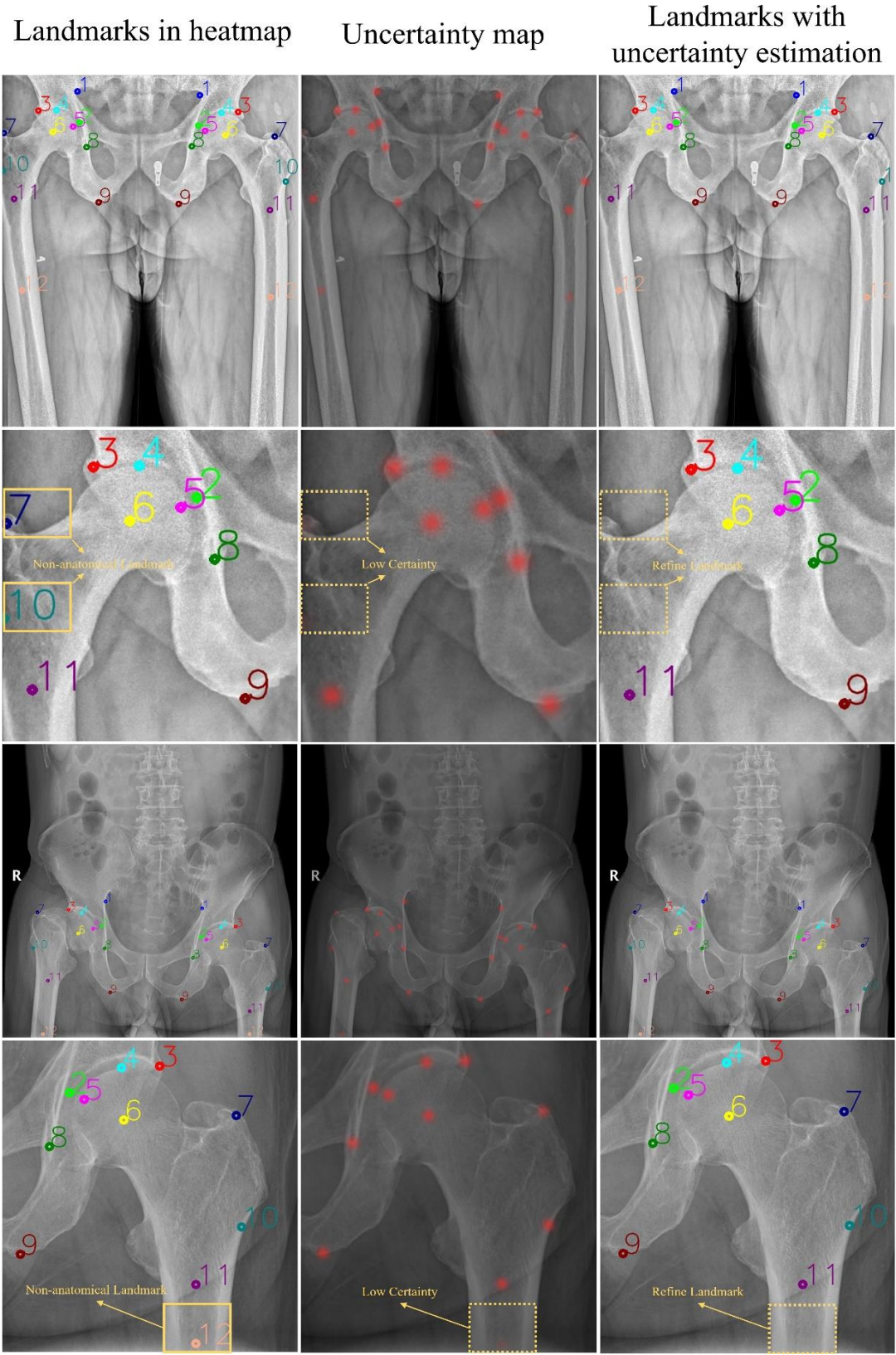

Figure 9 the uncertainty map produced by our UE module is presented, where brighter regions denote higher model confidence. Landmarks in heatmap in Figure displays the landmarks original predicted by the heatmap, and Landmarks with uncertainty estimation in Figure shows the refined landmarks after applying UE module. Landmarks enclosed in yellow boxes indicate model predictions that are not anatomically meaningful.

## 5. Discussion

In this study, the proposed method has achieved promising results in unstructured hip CT images. The Adaptive Wing Loss function effectively balances positive and negative sample representation in the heatmaps, enhancing the model's sensitivity to both and improving landmark detection accuracy.

Additionally, the SRF module significantly strengthens the network's ability to capture spatial relationships between landmarks. Ablation studies indicate that incorporating the SRF module into HRNet reduces NME, MRE by over 40% across the entire validation set, underscoring its effectiveness in managing complex unstructured radiographic data.

Moreover, comparative experiments across multiple architectures validated the efficacy of the UE module. In UNSCT-HRNet, the UE module reduced NME and MRE by more than 60% in unstructured data, demonstrating enhanced robustness and providing a reliable measure of confidence in the model's predictions.

# *6. Conclusion*

This paper emphasizes the importance of study for unstructured hip CT images, exploring how to quantify the uncertainty of landmark prediction while suppressing the output of non-anatomically landmarks. We proposed a method to detect anatomical landmarks from radiological images, which is able to model spatial and anatomical relationship features between landmarks, quantify the uncertainty of landmark predictions and suppress non- anatomically landmark outputs for accurate assessment of THA. By incorporating SRF module, UE module, and the HRNet architecture, our method achieved promising results on unstructured data.

# *Declaration of competing interest*

The authors declare that they have no known competing financial interests or personal relationships that could have appeared to influence the work reported in this paper.

# Acknowledgments

This research was Supported by the Fundamental Research Funds for the Central Universities, University of Electronic Science and Technology of China, Project No. Y030242063002050.

Special thanks to Professor Yutang Ye for his support and guidance throughout this study.

# Data Availability

The research data includes patient data and is available upon request.